\documentclass[letter,usenatbib]{mnras}
\pdfoutput=1

\usepackage{siunitx}
\usepackage{color}
\usepackage{natbib}
\usepackage{graphicx}
\usepackage{tabularx}

\font\FermiSmallfont=cmssq8 scaled 1200

\def\LANLppthead#1#2#3{
\null 
\begin{center}\vskip -1.0truein{\hbox to 7.5truein {
\hfill
\vbox to 1in {\vfill \FermiSmallfont
              \hbox{#1}
              \hbox{#2}
              \hbox{#3}
              \vfill}
}}\vskip-0.0truein\end{center}}

\def\ts{\rm TS}

\title[Blazar Variability from Wakefield Phenomena]{Observational Signatures of Gamma Rays from Bright Blazars
  and  Wakefield Theory}

\author[N.\ E.\ Canac et al.]{N.\ E.\ Canac$^{1}$,
  K.\ N.\ Abazajian$^{1}$, T.\ Tajima$^{1}$,
  T.\ Ebisuzaki$^{2}$ and S.\ Horiuchi$^{3}$
  \\
  $^{1}${Department of Physics and Astronomy, University of California, Irvine, Irvine, California 92697, USA}\\
  $^{2}${RIKEN, Wako, Saitama, Japan}\\
  $^{3}${Center   for Neutrino Physics, Department of Physics, Virginia Tech,  Blacksburg, VA 24061, USA}
}

\date{Accepted XXX. Received YYY; in original form ZZZ}

\begin{document}
\label{firstpage}
\pagerange{\pageref{firstpage}--\pageref{lastpage}}

\maketitle

\begin{abstract}
Gamma-ray observations have revealed strong variability in
blazar luminosities in the gamma-ray band over time scales as short as
minutes.  We show, for the first time, that the correlation of the
spectrum with intensity is consistent with the behavior of the
luminosity variation of blazar SEDs along a blazar sequence for low
synchrotron peak blazars.  We show that the observational signatures
of variability with flux are consistent with wakefield acceleration of
electrons initiated by instabilities in the blazar accretion
disk. This mechanism reproduces the observed time variations as short
as 100 seconds.  The wakefield mechanism also predicts a reduction of
the electron spectral index with increased gamma-ray luminosity, which
could be detected in higher energy observations well above the inverse
Compton peak.

\end{abstract}

\begin{keywords}
acceleration of particles -- plasmas -- galaxies:active -- gamma-rays:galaxies.
\end{keywords}

\section{Introduction}
Bright gamma-ray emitting blazars have been detected to have
variability in flux by a factor of two and more on time scales from
minutes to weeks
\citep{Albert:2007zd,Aharonian:2007ig,Aleksic:2011hr,Abdo:2010gba,Abdo:2010rw}. This
short-scale temporal variability has presented a strong challenge to
jet emission models. The spectral indices are often also time
variable, around the index of two or greater
\citep{Abdo:2010gba,Albert:2007zd,2013RAA....13....5C}. Furthermore,
an anti-correlation of the spectral index and the gamma-ray flux is
also reported \citep{Abdo:2010gba}, which implies a connection to the
underlying emission mechanism.  The High Energy Stereoscopic System
(HESS) telescope has observed flares on time scales of $\sim$200
seconds in Very High Energy (VHE) $>100\rm\ GeV$ gamma rays with peak
fluxes at a factor of two above the average flux toward the blazar PKS
2155-304~\citep{Aharonian:2007ig}.  The MAGIC telescope has seen flux
variability by an order of magnitude over several minutes, with a
hardening of the spectrum (corresponding to lower spectral index) with
increasing flux~\citep{Albert:2007zd}.  Similar variation is seen in
PKS 1222+21 in VHE observations by MAGIC, with a potential
smooth spectral connection to the GeV scale in observations by the
Large Area Telescope {\it Fermi Gamma Ray Space Telescope} (Fermi-LAT)
\citep{Aleksic:2011hr}.

The lack of a strong spectral break in the VHE gamma rays has
indicated that the jet emission mechanism must be outside of the
broad-line emission region (BLR), i.e., away from the region's very
high photon densities.  From the temporal variability it can be
inferred that the emission region must be very compact, of order
$\sim\! 10^{14}\rm\ cm$ from causality.  To avoid the optical depth
within the BLR region, it has been hypothesized that there are
small-scale regions embedded within a larger jet
\citep[e.g.,][]{Giannios:2009kh}.  Accommodating these phenomena in
traditional jet emission models has proved so challenging, as to
motivate new axion-like-particles to allow for photon-axion mixing to
potentially provide the temporal variability and spectral features
(e.g., \citealt{Tavecchio:2012um}). Alternatively, studies have tied
the timescales optical variability with episodic emission in the
magnetic field of the jet
\citep{Marscher:2008aa,Edelson:2013oaa,Edelson:2014xra}.

Fermi-LAT observations have revealed GeV gamma-ray signatures from the
brightest blazars \citep{Abdo:2010gba}.  Notably the gamma-ray
observations from Blazar AO 0235+164 and 3C 454.3 show
similar properties such as:
\begin{enumerate}
\item The photon index is around 2 or slightly above 2 in their
  respective lowest value;
\item the photon index varies rapidly from its lowest (around 2) to
  highest (around 2.8 or so) value over a period of several weeks;
\item the luminosity (flux) of the gamma rays also varies rapidly
  over the same time period of several weeks, by as much as a factor of five;
\item and, most importantly, the luminosity peak and the valley
  (hardening) of the photon index positively coincide. In other words,
  the time variations of the photon index and luminosity 
  faithfully anti-correlate. This anti-correlation clearly persists
  regardless of different periods from hundreds of seconds to months,
  demonstrating a remarkable universality in the phenomenon. 
\end{enumerate}
We show in this paper, for the first time, that the rapid variations
of high-energy gamma emissions and the strong anti-correlation between
the luminosity and photon index is consistent with a blazar SED
sequence shift. We further show that the timing of the variations is
consistent with magneto-rotational episodic instabilities, recently
studied in \citet{Mizuta:2017phr}. We show
that the inherent acceleration mechanism may be probed by further
analysis of the highest energy blazar spectra and their variability. 

The above indication of the position away from the high density
region, the inferred compactness of the emission region, and the
minute time-scale variabilities, all indicate a mechanism of the
variability arising from high-energy electrons from more compact and
robust energy conversion than can be provided by Fermi stochastic
acceleration. The wakefield acceleration of electrons can provide such
a viable mechanism with the properties that may fit and explain these
features \citep{Ebisuzaki:2013lya}. When the magneto-rotational
instability enhances magnetic fields in the inner accretion disk of
the blazar \citep{Balbus:1991ay,Matsumoto1995} and episodically causes
a large variability in matter accretion, this can severely disturb the
base of the jets. These shock waves at the jet base, {\it i.e.} large
amplitude Alfven waves, propagate along the jets, eventually
mode-convert themselves into intense relativistic electromagnetic (EM)
pulses along the jets. These excitations of EM pulses are capable of
giving rise to bow wakes, which accelerate electrons to high energies
(only limited by their emissions of radiation)
\citep{Ebisuzaki:2013lya}. This process is prompt, energy-efficient
and robust, and known to hold a stiff energy power spectrum.

The additional features of blazar’s observational phenomena include
that high energy gamma photon emissions are episodic: They contain
very short time-scale structure within one burst, as well as longer
evolution and much longer-scale variability. Because of these
features, we can relate the above observed properties to their
physical origins. The central compact object black hole is the source
of the energy of the burst, which is accompanied by an active
accretion disk, which in turn spawns out a pair of (spiral)
jets. Young objects of AGN/galactic systems provide active dynamical,
rather than stochastic, and robust plasma and magnetic activities (as
described in \citet{Tajima2002plap.book.....T}). The directionality of
blazar and good robust conversion of gravitational energy of such a
disk into the accretion and emission processes are required to be
explained in any complete model. In order for such active and large
energetic occurrence to happen, we surmise that ``collective'' (as
opposed to ``individual'') forces convert energy of gravitation
effectively and in a short matter of time. In prior work, it was shown
that short-range collisional processes are likely too slow and
non-dynamical, while the plasma’s collective interaction is
long-ranged and far-reached \citep{Ebisuzaki:2013lya,Mizuta:2017phr}.

Our model for the gamma ray variability from blazars is as follows
(see also \citet{Ebisuzaki:2013lya}): The gravitational energy in
young accretion disk rotational motion is stored as magnetic energy
buildup by the shear rotation of the disk via the magneto-rotational
instability (MRI) \citep{Balbus:1991ay}.  The buildup may be disrupted
due to its explosive growth \citep{Mizuta:2017phr}.  This eruption of
the magnetic fields in the disk accompanies disruptive episodic large
accretion of disk matter toward the central object as well as its
jets.  Severe shaking of the base of the jets of AGN---jets
accompanying spiral magnetic fields---causes severe Alfvenic shocks
whose wavelength along the jet. This is determined by the size of the
infalling accretion material (as in \citet{Mizuta:2017phr}). The shock
wave propagation along the jets from their source converts modes of
Alfven waves eventually into electromagnetic (EM) pulses, as the
density and magnetic field strengths go down as they propagate, with
its frequency preserved. This constitutes a spontaneously generated
large-scale emission of wakefields, and gives rise to immense
wakefield acceleration of leptonic and hadronic cosmic rays. The sum
total of these process can give rise to the observed gamma ray
variability, and we show below that the observations exhibit many
features that are consistent with this model’s consequences.

In Section 2, we review the theory of wakefield acceleration, and in 
Section 3, we review the gamma-ray emission processes relevant
for the wakefield acceleration mechanism in the context of blazars. 
In Section 4, we detail our gamma-ray analysis setup, and present
our results in Section 5. We conclude in Section 6. 

\section{ Magneto-rotational Instability and Wakefield Acceleration}

As a theory for the underlying electron energy injection,
ponderomotive acceleration, a version of wakefield acceleration
\citep{1979PhRvL..43..267T}, provides a theoretical framework of the
extremely relativistic collective acceleration mechanism in an
idealized case \citep{Ebisuzaki:2013lya,Ebisuzaki:2014vda}. In this
mechanism, the wave front is regarded as one-dimensional only
depending on the coordinate in the direction of the Alfven shock---and
its mode-converted electromagnetic (EM)---waves along the jet
propagation. In this one-dimensional model the coherence of high
energy acceleration is guaranteed with the asymptotically tending
velocity of the EM group velocity being the speed of light $c$
\citep{1979PhRvL..43..267T,Ashour-Abdalla:1981}.  This is best
realized when the pulse contains a single frequency carrier EM wave,
just as is the case of a laboratory laser experiment.  This may not be
necessarily the case in our astrophysical setting, in which we expect
a multiple set of frequencies of EM waves.  However, we note that the
group velocities of various EM waves with different frequencies are
nearly equal to $c$ in the one-dimensional case. Furthermore, as
explained in the theory of \citet{Ebisuzaki:2013lya}, the pulse is
generated by the striking of an acceleration of matter ejected by a
major disruption of the AGN accretion disk---the so-called
magneto-rotational instability (MRI). As such, an eruption could
be represented by one major outburst (though could be a series of such)
which results in a predominance of a single pulse with a typical
length of the disk thickness.  This results in the following
situation: while the accelerating field acquires a complex phase
structure, the phase velocity of each portion of the wave substructure
is again close to $c$, i.e., $c \sqrt{(1- \omega_p^2 / \omega(z)^2
  )}$, where $\omega(z)$ is the frequency of that local, where $z$ is
along the jet propagation and substructure's frequency and $\omega_p$
is the plasma frequency where the pulse propagates. This mechanism is
one of the origins of relativistic coherence, as pointed out in
\citet{Tajima:2010}.

The intense EM pulse is produced via collective (non-collisional) and
coherent robust acceleration called the wakefield process
\citep{1979PhRvL..43..267T}. This acceleration mechanism is via the
ponderomotive force that is from the Lorentz term $q(v\times B )/ c$.
Suppose that the EM pulse is propagating in its axial (along the jet)
direction (all it the $z$-direction) with the electric polarization of
the EM pulse in the $x$- and the magnetic in the $y$-direction
(circularly polarized case can be easily accommodated to the same
effect).  The rapidly oscillating EM fields cause a charged particle
(such as an electron) to execute a figure-eight oscillatory orbit to
relativistic velocities so that $ v \sim c $ in the $x$-direction by
this electric field $E$; Meanwhile, the magnetic field $B$ in the
$y$-direction causes this particle accelerated in the $z$-direction
(longitudinal direction, the same as the EM pulse propagation) via the
$ v \times B $ force, which is called the ponderomotive force
\citep{Tajima:1985}. In less extreme laboratory cases the
ponderomotive force is often calculated by considering the envelope of
the EM pulse \citep{Sprangle:1988}. However, in our present
ultrarelativistic regime (we define this as $ q E / m \omega c = a_0
\gg 1 $, where $E$ is the electric field of the EM pulse, $m$ the mass
of the electron, $\omega$ the frequency of the EM (and Alfven) wave,
and $ a_0 $ the normalized vector potential of the EM pulse (the
strength parameter of the EM wave)), the ponderomotive force need not
be averaged over the pulse length, but rather its raw Lorentz force
may directly impact acceleration in the longitudinal direction.  Thus,
noting that the magnetic field $ B \sim E $, in the present underdense
jet region \citep{Ebisuzaki:2013lya}, and $ v_y \sim c $, we obtain
the strength of the ponderomotive force $ F_p $ along the longitudinal
direction is $ F_p = m c \omega a_0 $.  This ponderomotive force
causes charge separation and triggers large plasma wave in the
longitudinal direction with the accelerating field of $E_{TD} = m c
\omega_p a_0 / e$. The field is called the Tajima-Dawson field that is
characteristic accelerating field of the wakefield acceleration,
\citet{1979PhRvL..43..267T}, where $e$ is the electron charge and
$\omega_p$ is the plasma frequency.  The plasma oscillations thus
excited having the phase velocity of speed of light $c$ are so far out
of the plasma electron thermal velocity  $v_{th}$ that the plasma’s
disturbance is known to not have a disturbing influence on the
wakefields. In addition, since in our case the EM pulse is in the
ultrarelativistic regime, there is another mechanism to
stabilize the wakefields.  This is because in the ultrarelativistic
regime the relativistic coherence plays additional role to stabilize
the acceleration process \citep{Tajima:2010}.  If the jet is flowing
with a relativistic Lorentz factor $\Gamma$, the ponderomotive force
becomes $ F_p = \Gamma m c \omega a_0 $.

While the relativistic coherence is preserved in the one-dimensional
situation, the escape of particles in this multi-frequency EM
drive becomes much more frequent than that of a single carrier case.
This incessant de-trapping (and subsequent re-trapping and its
repetition of these processes) gives rise to the emergence of
phase-induced stochasticity \citep{Mima1991}. According to this
theory, the resultant energy spectrum in one-dimension takes $E^{-2}$, 
{\it i.e.}, the energy spectral index of 2.  That is, in the most ideal
situation of purely one-dimensional ponderomotive acceleration, the
energy spectrum takes index of 2.  In less idealistic cases of
two-dimensions (or three-dimensions), the ponderomotive acceleration
incurred by waves that point to various directions (albeit within a
narrow cone around the jet direction $z$) now makes the de-trapping
more rapid so that the accelerating length per one episode of the wave
trapping of particles becomes shortened and the energy gain less. We
call this as the shortening of the de-phasing length as a function of
the dimensions of the wave structure \citep{1979PhRvL..43..267T}. This
leads to an energy spectral index greater than 2, with more particles
dominating in lower energy bracket.  If we compare the first
one-dimensional case with the two-dimensional (three-dimensions) case,
the amount of energy gain is higher in one-dimension than in
two-dimensions, as more coherent energy gain is realized in the
former. Thus it is this intrinsic mechanism of the ponderomotive
acceleration that makes the lower index case acquire greater and more
coherent energy gain in one-dimension than the cases in more spread
wave propagation in two-dimensional (and three-dimensional)
cases. Therefore, in ponderomotive acceleration, the energy
spectral index naturally anti-correlates with the particle energy gain
(and therefore the luminosity when converted into gamma rays).

On the other hand, the recent wakefield acceleration mechanism
\citep{Ebisuzaki:2013lya} has been suggested of the genesis of highest
energy cosmic rays accelerated via wakefields generated by the Alfven
shock emanated from the jet from an AGN accretion disk disruptions.
It has an embedded feature of accelerating both protons (ions) as well
as electrons simultaneously.  Moreover, it has built-in
characteristics of the following:  While the Fermi mechanism is
fundamentally stochastic \citep{1954ApJ...119....1F}, the current
mechanism is based on the coherent baseline process with the
relativistic coherence, though there are elements that bring in
stochastic processes that overlay the coherent mechanism.  High-energy
electrons accelerated by this mechanism in its purest form (dominated
by nearly one dimensional collimation along the axis of the blazar)
have the power-law spectrum of energy with photon index of 2.  If
there are less ideal or less robust regime of its operation, the power
index would rise above 2.  When the wakefield generation is most
robust, naturally the luminosity is also highest.  The acceleration
process has inherent rapid time scales.  The shortest time variation
is about 100 seconds, reflecting the Alfven wave structure, the next
time hierarchy has days-weeks associated with the occurrence of the
accretion disruption interval, and the longest time scale corresponds
to the acceleration time of the highest energy cosmic rays ($1-10^3$
years).  This last time scale is primarily for protons.  For electrons
with much lighter mass, the acceleration time scale may be in the
range from 100 seconds to weeks, in other words, the first and second
time scales, though the time structure depends on the detailed
acceleration configuration.  Although the gamma-ray luminosity
integrated over $4\pi$ is on the order of $10^{41}\rm\ erg\ s^{-1}$
for a black hole of $10^8\ M_\odot$ and normalized accretion rate of
$0.1 M_\odot /\mathrm{yr}$, the
boosted-apparent luminosity reaches $10^{44}-10^{45}\rm\ erg\ s^{-1}$,
depending on the $\gamma$-factor of the jet.  This is within a few
orders of magnitude of the observed gamma-ray luminosities of
blazars. For example, blazar 3C 454.3 has a bolometric luminosity from
0.1 to 100 GeV of $4.1\times 10^{48}\ \mathrm{erg\ s^{-1}}$
\citep{Nolan2012}. The episodic dynamics of the magnetic eruption of a
black hole's accretion disk, along with the associated intense
disruption of their jets was studied via three-dimensional
general-relativistic magnetohydrodynamics in \citet{Mizuta:2017phr}.

\begin{figure}
\includegraphics[width=\columnwidth]{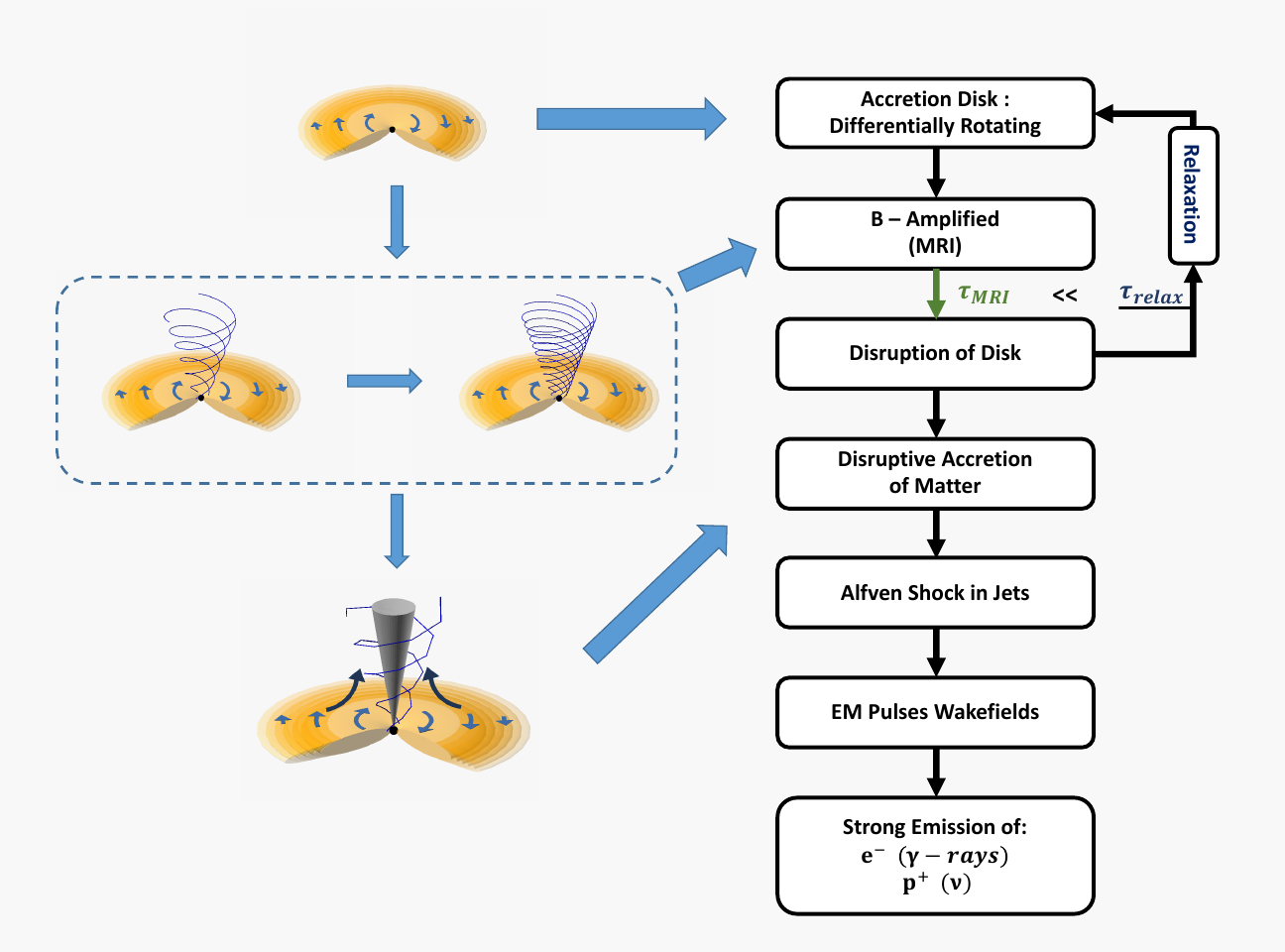}
\caption{The accretion disk that contains embedded magnetic fields does two
effects: (1) it will stretch the embedded magnetic fields by the
differential rotation of the accretion disk, which tends to amplify
the magnetic fields; (2) Such growth of magnetic fields and the
differential rotation lead to the MRI
\citep{Balbus:1991ay}.  The MRI triggered disk
instability leads to the accretion of concentrations of matter toward the center
of the disk and therefore the black hole.  These concentrations of
accreting matter should run into the feet emanating magnetic fields
and jets out of the central black hole. These collisions of the plasma
concentrations into the feet of the jets will disturb the jets’ base. Once the
magnetic fields give rise to the burst of matter to accrete, the
magnetic fields leave the disk and relax to the state in which little
magnetic fields in the disk. The disk repeats this pattern, as shown
in the flowchart.}
\label{fig:flowchart}
\end{figure}

We give a flowchart timescale of the flux variation as connected
between our MRI model and our observations in Fig.~\ref{fig:flowchart}.
The accretion of plasma amounts to intense excitation of pulsed Alfven
shocks at the feet of the jets. A diagram of this mechanism is shown
in Fig.~1 of \citet{Ebisuzaki:2014vda}.  These shock waves propagate
along the jets, turning themselves into EM pulses of super high
intensity, as the plasma density in the jet make the Alfven wave
converting itself into EM pulse with phase velocity at speed of light
$c$.  The likely intensity of these pulses that are characterized by
the normalized vector potential, $a_0 = eE_0 / m c w_0$, where $E_0$
and $w_0$ are the EM pulse’s amplitude and frequency, of the EM pulses
are far beyond unity.  Such intensity is termed as ``relativistically
intense'' pulses, as the EM acceleration of particles reach to
relativistic momentum in a single period of its EM pulse. These
relativistically intense (coherent) EM pulses are known to produce
intense longitudinal (i.e. in the direction of the EM pulse
propagation, not transverse to that direction) acceleration of charged
particles (both electrons in its negative phase and positrons and
protons in its positive phase) via the ponderomotive potential of
these intense EM pulses. This acceleration process has been known and
called as wakefield acceleration \citep{1979PhRvL..43..267T} and its
energy gain $E_\text{acc}$ is theoretically and experimentally well
known as $E_\text{acc} = mc^2 a_0^2 n_\text{cr} / n$, where the
quantities $n$ and $n_\text{cr}$ are the plasma density and the
critical density\citep{2017NCimR..40...33T}.

\begin{figure}
\includegraphics[width=\columnwidth]{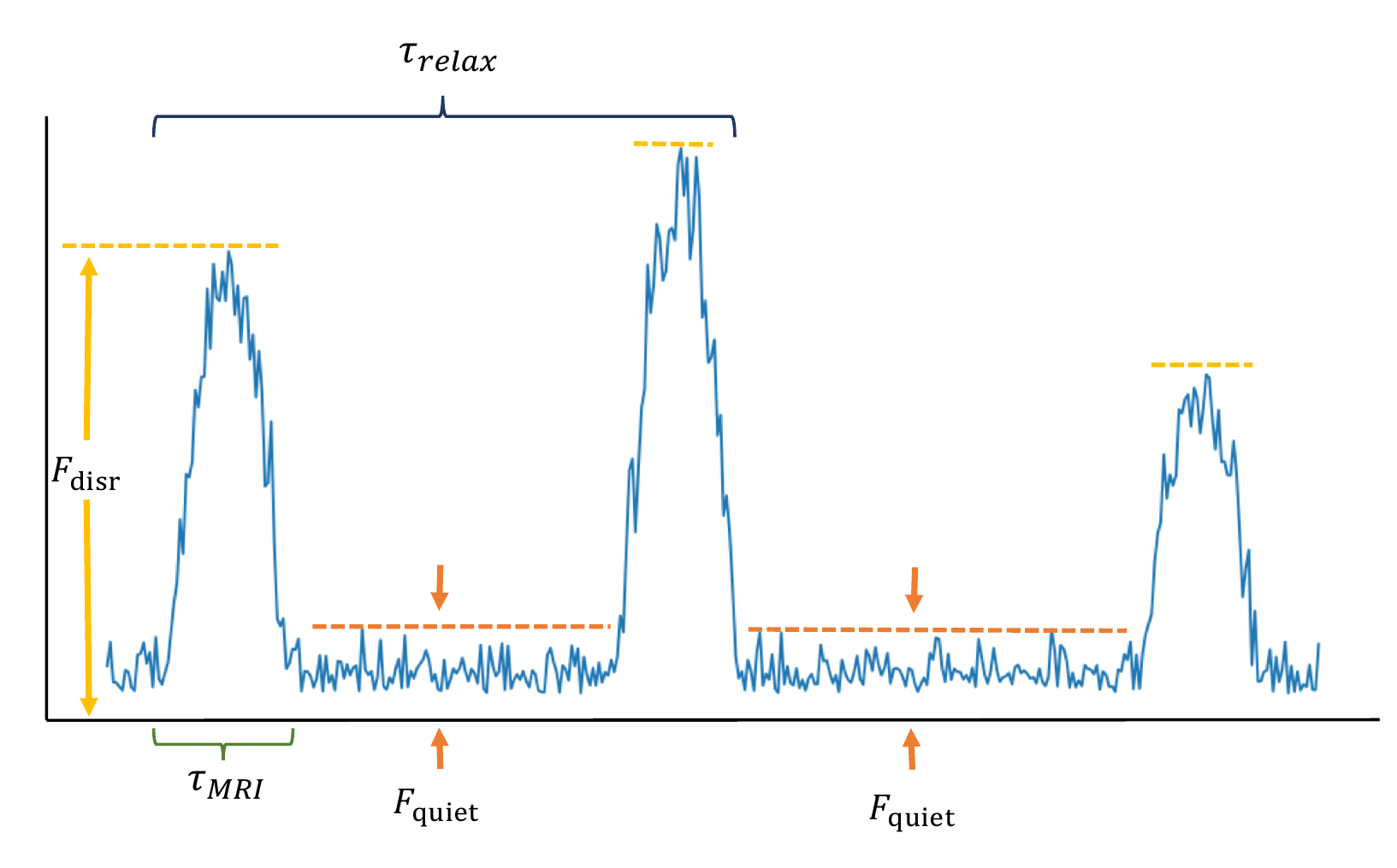}
\caption{This is a cartoon depiction of the MRI instability versus
  relaxation timescale. The MRI excitation timescale is
  $\tau_\text{MRI}$, originally calculated by \citet{Balbus:1991ay},
  and the relaxation and spin-up scale timescale $\tau_\text{relax}$
  is given by \citet{Ebisuzaki:2014vda} to be $\tau_\text{relax} \sim
  10 \tau_\text{MRI}$. The lower flux level ($F_\text{quiet}$) is the
  quiet phase flux as given by such theory as in
  \citet{1977MNRAS.179..433B}, while the high flux phase,
  ($F_\text{disr}$) is given by the disruptive accretion phase such as
  by \citet{Mizuta:2017phr}.}
\label{fig:episodic}
\end{figure}

Because of the genesis of the intense accelerating field generated by
the above process characterized by the parameter $a_0$ (which can be
looked upon as the relativistic Lorentz factor of the EM pulse), we
understand that whenever the MRI is incurred in the disk, the
accretion of plasma concentrations toward the black hole happens, MRI triggers
the above pulse generation in the jets and thereby both electrons and
positively charged particles accelerated by the above intense pulse
are emitted synchronously right after this accretion event.  Because
the MRI is episodically excited (Fig.~\ref{fig:flowchart}), the emission
of electron acceleration and ions is also excited, the former of which
may be observed as increases in gamma ray flux, while high energy
protons (and perhaps neutrinos) may also be emanated.  We show this
episodic flux and timescale relationship, and its relation to our
observations, in Fig.~\ref{fig:episodic}. As we discuss above, the
higher flux periods are correlated with higher electron spectral
index, but that is modulated by the synchrotron self-compton model, as
discussed below. 

The theoretical prediction also indicates that the larger the central
mass of the black hole is, the longer the repetitive recursion period
of the MRI.  Therefor, in future work, one can assess the mass of the
central black hole from the repetitive period observed by the gamma
rays.

\section{Synchrotron to Inverse Compton Acceleration}

\begin{figure}
\includegraphics[width=\columnwidth]{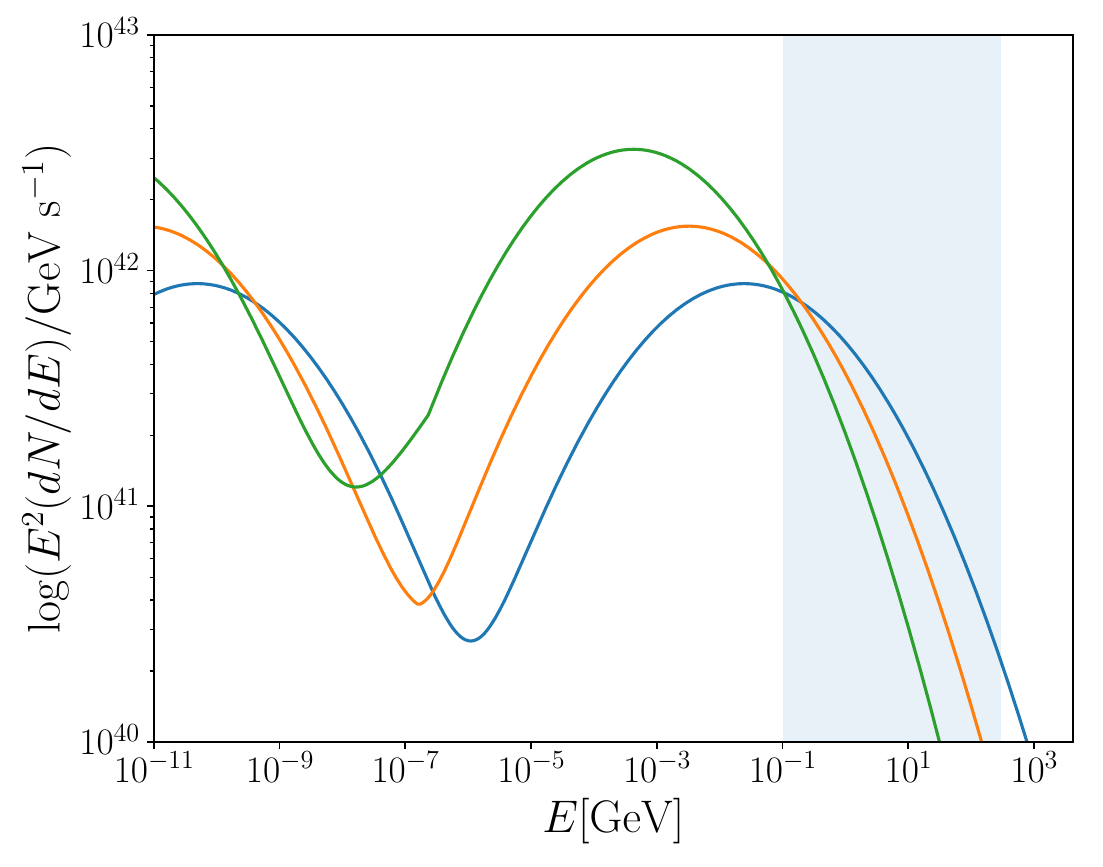}
\caption{Shown is a standard blazar sequence model for a typical
  blazar SED, based on \citet{inoue:2008pk}. The energy band of our
  observations is indicated by the filled blue box.  As can be seen in
  this illustration, a higher luminosity (green curve), relative to a
  lower luminosity (blue curve), would correspond to a photon spectral
  index much steeper than $\Gamma=2$ within the observational window.
  This is what we dub the ``internal blazar sequence.''  A spectral
  index of $\Gamma=2$ is horizontal on this plot. The orange curve
  represents the possibility of observing the photon spectrum that is
  more closely-tied to the intrinsic electron spectrum at higher
  energies than the inverse Compton peak intensity.}
\label{fig:BlazarSpectrum}
\end{figure}

In this section, we briefly review how accelerated electrons lead to
the observed gamma-ray spectrum. For the majority of blazars
the gamma-ray emission can be well described by the process of synchrotron
self-Compton \citep{2011hea..book.....L}. This is the process by which
synchrotron photons, produced in great abundance by high-energy
electrons in the magnetic field of the blazar jet, are up-scattered by
inverse-Compton collisions by these same electrons. Photons emitted by
the accretion disk may also be up-scattered. The intensity spectrum
has a characteristic double-humped feature where the 
high-energy gamma-ray emission is an inverse-Compton processed
up-scattering reflection of the lower-energy emission
\citep{fossati:1997vu,fossati:1998zn,donato:2001ge}. A model spectrum
of this variety is shown in Fig.~\ref{fig:BlazarSpectrum}. Here we
review the basic physics of the synchrotron mechanism that feeds into
the synchrotron self-Compton process, based on \citet{2011hea..book.....L}.
The high-energy gamma-ray spectrum of blazars may also have contributions from
high energy hadronic cosmic rays \citep{Takami:2013gfa}, though synchrotron
self-Compton models in general do a superior job in explaining the
broad blazar SED.

The primary energy flux at a given frequency $\nu$, $J(\nu)$, peaks in
synchrotron emission at a characteristic frequency,
\begin{equation}
\nu_c \approx \gamma^2 \nu_g = \left(\frac{E^2}{m_e c^2}\right)^2 \nu_g \, ,
\end{equation}
for an electron of energy $E$, where the gyrofrequency is $\nu_g =
eB/2\pi m_e$. The energy radiated in frequency interval
$(\nu,\nu+d\nu)$ is,
\begin{equation}
J(\nu) d\nu = \left(-\frac{dE}{dt}\right) N(E)\, dE \, ,
\end{equation}
where the electron population has a number distribution as $N(E) dE =
\kappa E^{-p}dE$. The energy loss rate for synchrotron radiation is,
\begin{equation}
-\left(\frac{dE}{dt}\right) = \frac{4}{3} \sigma_T c
\left(\frac{E}{m_e c^2}\right)^2 \frac{B^2}{2\mu_0}\, ,
\end{equation}
and the energy flux is,
\begin{equation}
J(\nu) \propto \kappa B^{(p+1)/2}\nu^{-(p-1)/2} \, .
\end{equation}
So, the emitted synchrotron spectrum $J(\nu)\propto \nu^{-a}$ is related to the intrinsic
electron energy spectrum $p$ as $a = (p-1)/2$. However, the
synchrotron emission undergoes self-absorption that alters the
observed spectrum. This is due to the limitation that
no region can emit incoherent radiation at an
intensity greater than that of a black-body at its thermodynamic
temperature. At low enough frequencies, the ``brightness temperature''
of the radiation approaches the ``thermal'' temperature of the
radiating electrons. The {\it brightness temperature} is derived from
that of a black-body, but is applicable to any emission process, and
is defined as $T_b = (\lambda^2/2k)(S_\nu/\Omega)$, where $\lambda$ is
the emission wavelength, $S_\nu$ is the flux density, and $\Omega$ is
the solid angle the source subtends at the observer. The effective
temperature of the electrons must match the brightness temperature,
and therefore the observed intensity must be \citep{2011hea..book.....L},
\begin{equation}
S_\nu \propto \frac{\nu^{5/2}}{B^{1/2}} \, .
\end{equation}
Therefore, a pure synchrotron self-absorption spectrum is a rising
spectrum as $I\propto \nu^{5/2}$ and then falls over to the inherent
synchrotron spectrum $I\propto \nu^{-(p-1)/2}$. The inverse Compton
spectrum reflects that since in that case the energy of the
up-scattered photons are $\hbar \nu = (4/3)\gamma^2\hbar \nu_0$, where
$\nu_0$ is the originating photon's frequency. 

There is one basic conclusion regarding the relation of the observed
photon spectrum in gamma rays or radio frequencies: the observed
energy spectrum could reflect the originating spectrum well above the
synchrotron peak where $I_\nu \propto \nu^{-(p-1)/2}$, but that is not
necessarily achieved in the given observational window, which could
lie below the synchrotron peak, near it, or above it, but still in its
spectrally curved region.

For an example blazar temporal sequence, we adopt the SED spectral
model first quantified in \citet{inoue:2008pk}. However, any smoothly
curved spectrum is likely sufficient, and we make no effort to fit a
model.  We show for the first time that, for the case of
low-synchrotron peak blazars we observe, the temporal variation of
flux is consistent with a given blazar shifting in flux and bolometric
luminosity along such a temporal blazar sequence. Our results of
observed gamma-ray spectra appear to be near the intensity peak
(overturn) of the inverse Compton emission, where $a=2$, and so are
not likely reflective of the source electron spectra. Higher energy
observations, such as with the High Altitude Water Cherenkov (HAWC)
gamma-ray observatory \citep{Lauer:2015bza}, MAGIC
\citep{Aleksic:2011bx}, H.E.S.S. \citep{Hinton:2004eu}, and,
eventually, the Cherenkov Telescope Array (CTA)
\citep{Acharya:2017ttl}, as well as temporal observations, may say
more about the intrinsic electron spectra. The cutoff shape could also
be due to the emergence of an upper electron energy for inverse
Compton emission, and that would have to be taken into account in
modeling of the highest energy spectra in future work
\citep{2012ApJ...753..176L}.

In summary, our observations could connect the temporal properties of
the wakefield acceleration mechanism, but a connection with the
intrinsic electron spectra is not possible with the lower-energy
Fermi-LAT spectra.  Importantly, the high-energy spectra will be
modified by intergalactic opacity, which becomes significant for
gamma-ray energies of more than 10 to 100 GeV
\citep{abdo:2010kz}. Therefore, any deconvolution of the electron
energies from the photon spectrum at high energies will have to take
into account effects of attenuation of gamma-rays as well for the
blazar observed \citep{Gilmore:2011ks}.

\section{Method}

Throughout our analysis, we use Fermi Tools version {\tt v9r33p0} to
study Fermi LAT Pass 7 reprocessed data taken from August 2008 to
February 2015 (approximately 85 months of data), using both front and
back-converting, {\tt SOURCE}-class photons. We select the twelve
blazars with the highest photon flux from the Fermi LAT second AGN
catalog \citep{ackermann:2011bg}. (Note that the analysis does depend
on the Pass version of the data.)  These blazars are listed in
Table~\ref{blazarlist}, along with their optical and SED class. For
each blazar, gamma rays within a circular region of interest (ROI), 7
degrees in radius and centered on the blazar, are selected, as was
done in \cite{Abdo:2010gba}. We use photons with energies between 100
MeV and 300 GeV, and apply the standard cuts recommended by the Fermi
collaboration to ensure data quality (zenith angle $< 100$ degrees,
\verb|DATA_QUAL = 1|, \verb|LAT_CONFIG = 1|).

\begin{table*}
\caption{List of blazars and their properties, sorted in order of 
decreasing photon flux, with their Pearson correlation coefficient 
between their photon flux and spectral index, along with the 
corresponding approximate p-value for the 50 time bin analysis.}
\label{blazarlist}
\begin{tabularx}{\textwidth}{l|cccccc}
  \hline\hline
Blazar name & Optical class & SED class & r & p-value & $\sigma^{2}_{NXS,flux}$ & 
$\sigma^{2}_{NXS,index} \times 10^{-2}$\\
\hline \\
3C 454.3 & FSRQ & LSP & -0.416 & \num[round-precision=2, round-mode=figures, 
scientific-notation=true]{0.00265222801623} & $2.44 \pm 0.010$ & $1.44 
\pm 0.14$ \\
PKS 1510-08 & FSRQ & LSP &  -0.441 & \num[round-precision=2, round-mode=figures, 
scientific-notation=true]{0.00134692630665} & $0.99 \pm 0.012$ & $0.04 
\pm 0.03$ \\
PKS 1502+106 & FSRQ & LSP & -0.681 & \num[round-precision=2, round-mode=figures, 
scientific-notation=true]{5.1123610085e-08} & $1.306 \pm 0.026$ & $0.42 
\pm 0.40$ \\
PKS 0537-441 & BL Lac & LSP & -0.379 & \num[round-precision=2, round-mode=figures, 
scientific-notation=true]{0.00668059118636} & $0.381 \pm 0.013$ & $0.38 
\pm 0.09$ \\
4C +21.35 & FSRQ & LSP & -0.491 & \num[round-precision=2, round-mode=figures, 
scientific-notation=true]{0.000297149722824} & $1.28 \pm 0.017$ & $0.68 
\pm 0.12$ \\
PKS 0426-380 & BL Lac & LSP  & -0.254 & \num[round-precision=2, round-mode=figures, 
scientific-notation=true]{0.0752701986899} & $0.34 \pm 0.013$ & $0.50 \pm 
0.10$  \\
Mkn 421 & BL Lac & HSP & -0.065 & \num[round-precision=2, round-mode=figures, 
scientific-notation=true]{0.652630588652} & $0.165 \pm 0.008$ & $0.03 \pm 
0.03$ \\
3C 279 & FSRQ & LSP & -0.265 & \num[round-precision=2, round-mode=figures, 
scientific-notation=true]{0.0633125219748} & $0.82 \pm 0.016$ & $0.23 \pm 
0.07$\\
3C 66A & BL Lac & ISP & 0.104 & \num[round-precision=2, round-mode=figures, 
scientific-notation=true]{0.471799122011} & $0.21 \pm 0.016$ & $1.49 \pm 
0.04$ \\
PKS 2155-304 & BL Lac & HSP  & 0.680 & \num[round-precision=2, round-mode=figures, 
scientific-notation=true]{5.62951817114e-08} & $0.40 \pm 0.023$ & $0.52 
\pm 0.12$ \\
PKS 0454-234 & FSRQ & -- &-0.587& \num[round-precision=2, round-mode=figures, 
scientific-notation=true]{7.40803889538e-06} & $0.34 \pm 0.013$ & $0.49 
\pm 0.12$ \\
PKS 0727-11 & FSRQ & -- & -0.145 & \num[round-precision=2, round-mode=figures, 
scientific-notation=true]{0.313666577429} & $0.31 \pm 0.019$ & $1.20 \pm 
0.28$ \\
  \hline\hline
\end{tabularx}

\end{table*}

\begin{figure*}
\begin{center}
\includegraphics[width=6.8truein]{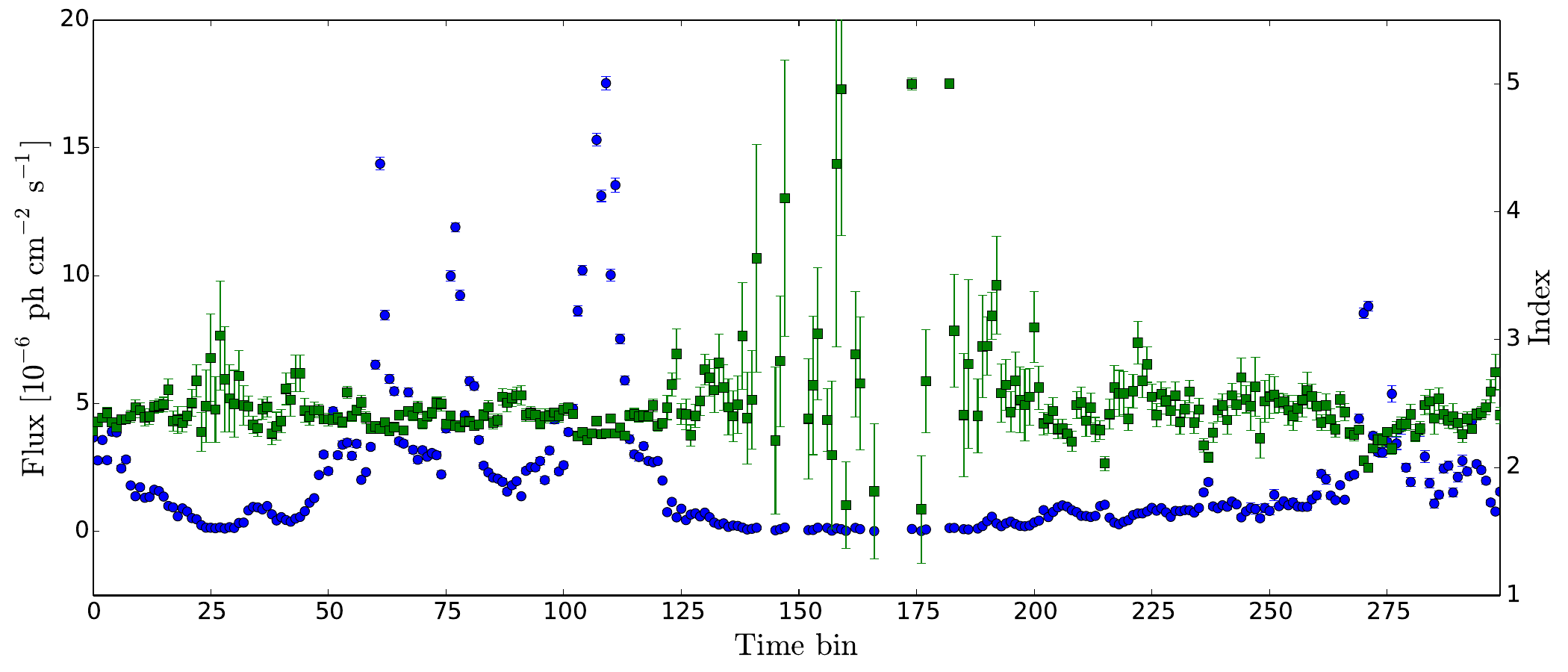}
\end{center}
\caption{Shown are the flux (blue circles, left axis) and spectral index (green 
squares, right axis) for 3C 454.3 in 300 time bins of 7.9 days duration.
An anti-correlation can be seen: the peaks in flux correspond to dips in 
the spectral index and vice versa.} \label{lightcurve}
\end{figure*}

Our model for each of the twelve regions of interest is composed of
all of the sources identified in the LAT 4-year point source (3FGL)
catalog, along with the recommended diffuse emission models associated
with the Galactic emission (\verb|gll_iem_v05_rev1|) and the isotropic
background (\verb|iso_source_v05|) which accounts for the
contributions from both the extragalactic background and cosmic ray
contamination.  Despite the fact that a number of blazars in this
analysis display curvature in their spectra over wide ranges in
energy, we employ an energy window where the curvature is minimal and
well-modeled by a power-law. We adopt simple power-law model
consisting of two free parameters, a normalization factor $N_{0}$ and
spectral index $\Gamma$, for the spectrum of each blazar.  The
convention used in this paper is that the spectral index $\Gamma$
should be taken to be positive, so that the photon flux is
proportional to $E^{-\Gamma}$, where $E$ is the energy. This is done
to provide a convenient means of characterizing the relative hardness
or softness of the photon spectrum, which is the primary goal of this
analysis.

Once we construct a model for the ROI around a particular blazar, we
use standard maximum likelihood methods to fit the free parameters of
the various gamma-ray sources in our model. To determine which
parameters to vary, we make use of the quantity $\ts$, which is
defined as twice the difference in log-likelihood between a model with
and without a particular source, i.e. $\ts =2\Delta \ln(\mathcal L)$
($\ts = 25$ corresponds to an approximate detection significance of
about $5 \sigma$ for point sources). We then determine the variability
of the photon flux and spectral index over time for each of the twelve
blazars by dividing up the full time range into time bins and
refitting the parameters for the source of interest, leaving all other
sources in the ROI fixed to their best fit values found from the full
time range. Because the blazar is, by a large factor, the brightest
source in the ROI for all cases, any variability of other sources is
minimal and would not alter our results. Our procedure, which makes
use of the LAT analysis scripts {\tt quickAnalysis}, {\tt quickLike},
and {\tt quickCurve}, is described in more detail below.

\begin{enumerate}
\item First, a binned likelihood analysis of the region is performed 
using photons from the full time range. This is referred to as the 
DC  analysis (analogous to ``direct current'').

\begin{itemize}
\item The raw photons file is filtered and processed according to the 
previously described specifications for each blazar using the 
{\tt quickAnalysis} tool.
\item The model file for the ROI is generated using the user 
contributed tool {\tt make3FGLxml.py} and then changing the spectral shape 
of the source of interest to a power law.
\item The parameters for sources with $\ts > 25$ are left free. Sources 
with $4 < \ts < 25$ have their spectrum fixed to their 3FGL values but 
their normalizations are left free. Finally, sources with $\ts < 4$ have 
all of their parameters fixed to their 3FGL values.
\item A standard binned maximum likelihood analysis is performed using 
the {\tt quickCurve} tool to find the best-fit values for all of the 
remaining free parameters in the model. This model is called the DC 
model.
\end{itemize}

\item Next, the full time range is divided up into time bins and a 
separate unbinned likelihood analysis is performed for each time bin. 
This is referred to as the AC analysis (analogous to ``alternating current'').

\begin{itemize}
\item The entire time range is divided into equally sized time bins of 
7.9 days, resulting in a total of 300 time bins. Though there may be 
temporal variation on time scales smaller than this, our timing 
resolution is limited by the detection significance of the source of 
interest, and we are unable to reliably probe time scales 
significantly shorter than about a week.
\item All of the relevant analysis files to perform an unbinned 
likelihood analysis for each time bin are generated using the 
{\tt quickCurve} tool, using the same specifications as in step 1.
\item The DC model from step 1 is copied, but all of the sources 
except for the source of interest are fixed to their best fit values 
from the DC analysis. Only the normalization factor and spectral index 
of the source of interest are left free.
\item An unbinned maximum likelihood analysis is performed in order to 
determine the best-fit values for the normalization and spectral index 
of the source of interest.
\end{itemize}

\item Step 2 is repeated once again, but using time bins that are 
larger by a factor of six (about 47.5 days), resulting in 50 time bins 
instead of 300.  This has the effect of increasing the detection 
significance of the source of interest in each time bin at the cost
of losing sensitivity to short time-scale variations. 
\end{enumerate}

\section{Results}

We find that all blazars displayed significant temporal variation in
both their flux and spectral index. Fig.~\ref{lightcurve} shows the
variation of the flux and spectral index for blazar 3C 454.3 over the
300 time bins. For nine out of the twelve blazars studied in this
analysis, we observe a weak to moderate anti-correlation between the
flux of the blazar and its spectral index.  Six of these are
statistically significant ($p$-value $< 0.05$).  In other words, for
most of the blazars, we observe the same ``harder when brighter'' effect
that has been noted in other analyses \citep{Abdo:2010gba}. This can
be seen in Fig.~\ref{lightcurve}, as the peaks in flux correspond to
dips in the spectral index and vice versa. 

\begin{figure*}
\begin{center}
\includegraphics[width=6.8truein]{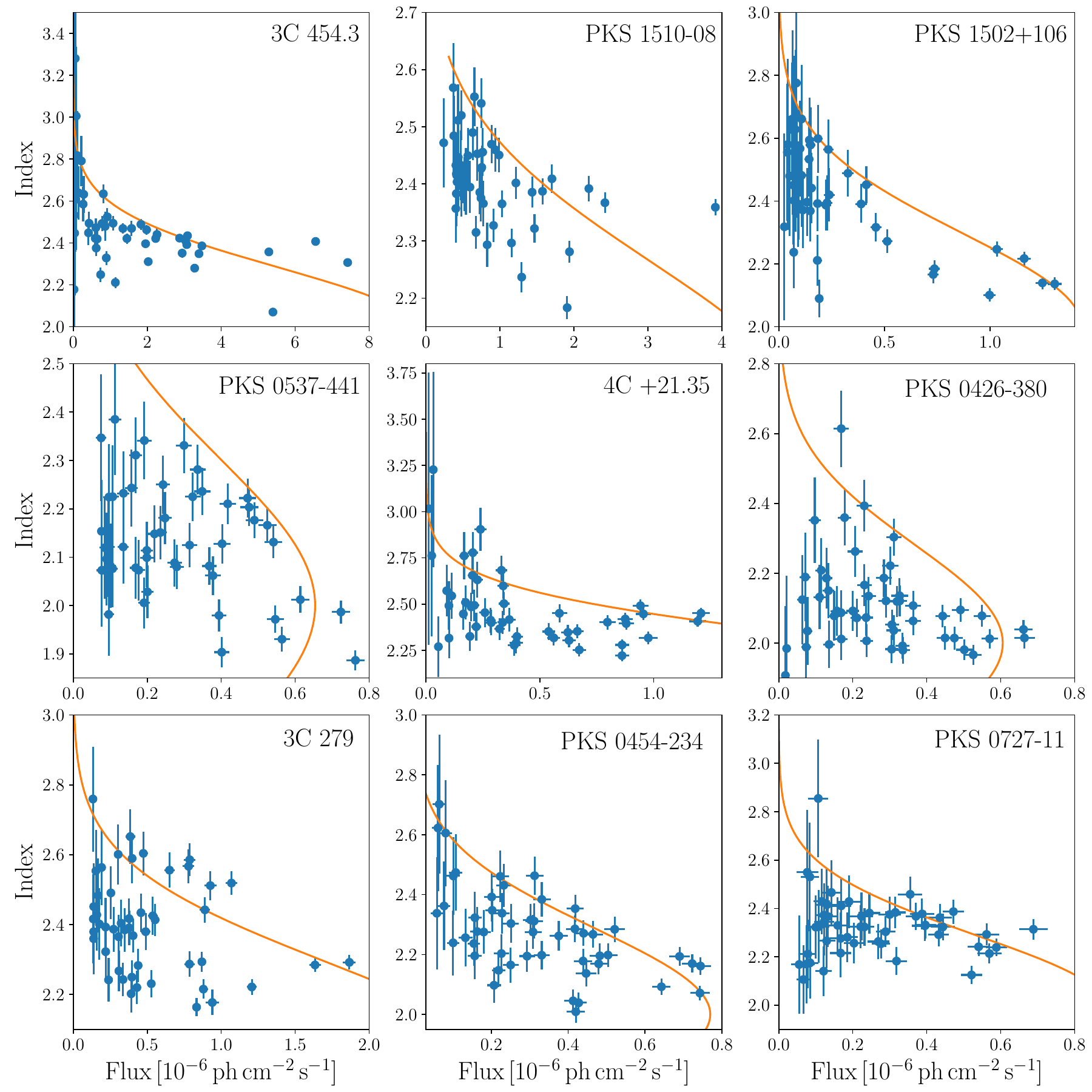}
\end{center}
\caption{Shown here are the flux vs.~index for the nine the blazars analyzed 
in the 50 time bin analysis that exhibit the flux-spectral
anti-correlation attributable to the inverse Compton up-scattering of
synchrotron photons. The orange curve is the slope from the blazar SED
model shown in Fig.~\ref{fig:BlazarSpectrum}, which is not fit to the
data, yet corresponds well with the observations. This is what we dub the
``internal blazar sequence.'' }
\label{flux_v_index_LSP}
\end{figure*}

\begin{figure*}
\begin{center}
\includegraphics[width=6.8truein]{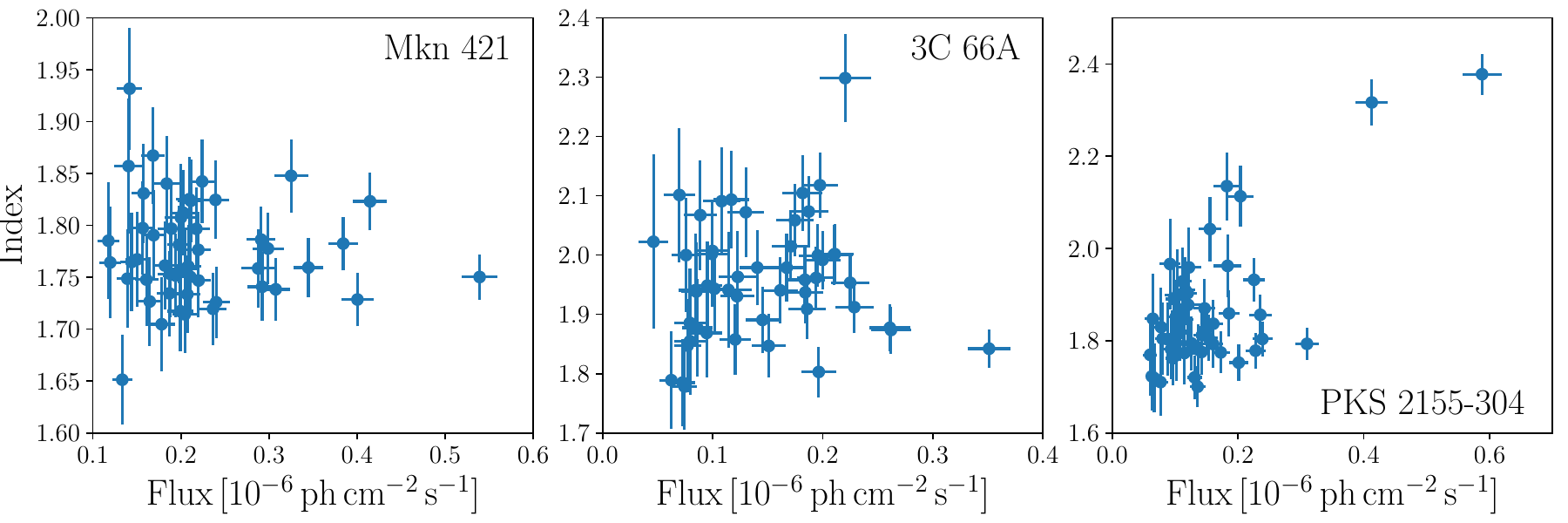}
\end{center}
\caption{Shown here are the flux vs.~index for three blazars not exhibiting
  the flux-spectral index anti-correlation. Unlike the LSPs in
  Fig.~\ref{flux_v_index_LSP}, two of the three, Mkn 421 and PKS
  2155-304, are HSP-BLLacs, and the third blazar, 3C 66A, is an ISP BL
  Lac. The pattern of FSRQs and LSP BL Lacs displaying the
  anti-correlation and HSP BL Lacs not exhibiting the effect was
  previously noted in \citet{Abdo:2010gba}. ISP-BLLacs tend to display
  it in some cases. We discuss these differentiated cases in
  Section \ref{discussion}.}
\label{flux_v_index_non}
\end{figure*}

This anti-correlation can be seen more clearly in
Fig.~\ref{flux_v_index_LSP} which shows the spectral index vs.~flux
for nine of the blazars in the 50 time bin analysis. The
anti-correlation is seen in both the 50 time bin analysis and the 300
time-bin analysis.  The Pearson correlation coefficients are shown in
Table~\ref{blazarlist} along with their corresponding approximate
$p$-values. Three blazars do not exhibit this anti-correlation
(Fig.~\ref{flux_v_index_non}). Two of the three, Mkn 421 and PKS
2155-304, are high synchrotron peak (HSP) BL Lacs, and the third
blazar, 3C 66A, is an intermediate synchrotron peak (ISP) BL Lac. This
same pattern of flat-spectrum radio quasars (FSRQs) and low
synchrotron peak (LSP) BL Lacs displaying the anti-correlation and HSP
BL Lacs not exhibiting the effect was previously noted in
\citet{Abdo:2010gba}, although they also note that ISP-BLLacs tend to
display it in some cases. Possible explanations for this behavior will
be elaborated on further in Section \ref{discussion}.

We estimate the intrinsic source variance by calculating the excess 
variance $\sigma^{2}_{XS}$, as described in \citet{Vaughan:2003by}. The 
excess variance is the variance after subtracting the contribution 
from measurement errors such as Poisson noise and is defined as,
\begin{equation}
\sigma^{2}_{XS} = S^{2} - \overline{\sigma^{2}_{err}} \, ,
\end{equation}
where $S^{2}$ is the sample variance of $N$ data points,
\begin{equation}
S^{2} = \frac{1}{N - 1} \sum_{i=2}^{N}(x_{i} - \overline{x})^{2} \, ,
\end{equation}
and $\overline{\sigma^{2}_{err}}$ is the mean square error,
\begin{equation}
\overline{\sigma^{2}_{err}} = \frac{1}{N} \sum_{i=1}^{N} 
\sigma^{2}_{err,i} \, .
\end{equation}
The normalized excess variance is given by $\sigma^{2}_{NXS} = 
\sigma^{2}_{XS}/\overline{x}^{2}$, and the error on $\sigma^{2}_{NXS}$ 
was calculated according to,
\begin{equation}
err(\sigma^{2}_{NXS}) = \sqrt{ \left(\sqrt{ \frac{2}{N} } \cdot 
\frac{\overline{\sigma^{2}_{err}}}{\overline{x}^{2}}\right)^{2} + 
\left(\sqrt{\frac{\overline{\sigma^{2}_{err}}}{N}} \cdot \frac{2 
F_{var}}{\overline{x}}\right)^{2} } \, ,
\end{equation}
as given in \citet{Vaughan:2003by}. Here, $F_{var}$ is the fractional 
root mean square (rms) variability amplitude and is simply the square 
root of the normalized excess variance, i.e., $F_{var} = 
\sqrt{\sigma^{2}_{NXS}}$. The measured values for normalized excess 
variance of the photon flux $\sigma^{2}_{NXS,flux}$ and for the 
spectral index $\sigma^{2}_{NXS,index}$ are also shown in 
Table~\ref{blazarlist}.

\section{Discussion}
\label{discussion}

We have established the anti-correlation between the
luminosity of gamma rays from blazars and their spectral index. The
anti-correlation is clearly consistent with a shift from off-peak to
peak portions of the blazar spectra, as shown in
Figs.~\ref{fig:BlazarSpectrum} and \ref{flux_v_index_LSP}, for the case
of LSP BL-Lacertae blazars. This is the first time that a
temporal-sequence model has been applied to show the origin of the
spectral-flux anti-correlation.

We followed in this study the earlier observations by
\citet{Abdo:2010gba} and \citet{Abdo:2010ge}. In our work, we scan
different blazars with different values of flux and index, as in these
prior analyses. When they surveyed different blazars, they showed
different rise times of the brightening and the lowering of the
spectral index as well as the interval times between such bursts of
luminosity surge. The rise (and fall) times in most cases scale with
the interval times between the events, i.e., the longer the rise time
of the burst is, the longer the interval is. This is predicted in the
magneto-rotational instability and wakefield acceleration model of
\citet{Mizuta:2017phr}.
Even though the time scale varies over
orders of magnitude, this phenomenon is universal. In fact,
\citet{Ebisuzaki:2013lya} predicts that all these events are
universally based on the wakefield excitation in the jets sharing the
common acceleration mechanism, which is the basis for this
universality. 

We note that the observed phenomenon of the anti-correlation between
the luminosity and the spectral index remains manifest, regardless of
the mentioned rise time and interval time scales.  According to the
theory of \citet{Ebisuzaki:2013lya}, the periods of both the rise time
and the interval time scale are proportional to the mass of the
central AGN mass \citep{Ebisuzaki:2013lya}. In other words, though an
AGN with different mass may show a proportional time scale of
variability, these phenomena are common and the luminosity-index
anti-correlation is one of the strongest evidence for the universal
nature of acceleration mechanism. It should be further noted that not
only these universal phenomena are expected and/or observed in
blazars, but also these should be expected for microquasars with much
smaller masses, as predicted in \citet{Ebisuzaki:2014vda}. Therefore,
it is encompassing several orders of a wide range of masses from
microquasars to most massive AGNs. From this discussion, we now
suggest the following prediction: from the period time scale of the
blazar gamma ray emissions, we can surmise the mass of the central
object of the particular blazar. Therefore, the intrinsic luminosity
of blazars, though collective of apparent luminosity depend on the
observational conditions, can also tell us about the mass of the
central object, as they also scale proportional to the mass
\citep{Ebisuzaki:2013lya,Ebisuzaki:2014vda}.

Our own analysis (Fig.~\ref{lightcurve}) and subsequent analyses also
underline the following picture: when the gamma-ray emission from
blazars increases, the gamma-ray spectrum tends to become harder. This
observed tendency, which is in agreement with previous observation
\citep{Abdo:2010gba}, is consistent with a temporal SED sequence,
shifting emission from near-peak to peak inverse Compton emission
flux.  Higher energy observations, such as with the HAWC observatory,
may be able to probe the intrinsic electron spectrum to test its
consistency with the wakefield acceleration electron spectra in the
jets.  For HSP and ISP blazars, we do not see the anti-correlation as
we had for LSP blazars with lower-energy synchrotron peaks. This is 
due to the fact that, for HSP and ISP, the spectral peak is within the
energy window not exhibiting the monotonic SED sequence
anti-correlation. In addition, several of our tested HSP and ISP
blazars exhibit no significant curvature in their spectra
\citep{abdo:2009iq}.

The episodic disk instability launches the energetic Alfven shocks,
and their subsequent electromagnetic pulsations, such as in the
Ebisuzaki-Tajima mode \citep{Ebisuzaki:2013lya}. This model naturally
points to a localized source \citep[as observed
  in][]{Abdo:2010gba,Albert:2007zd}. Such localized emission of
gamma-rays (arising from accelerated electrons) and expected localized
acceleration of ions from such locations are integral properties of
the wakefield acceleration mechanism. Furthermore, the wakefield
acceleration of ultra-high energy cosmic-ray ions has the unique
property of localized gamma-ray emission, while the Fermi acceleration
mechanism relies on a diffuse site and therein lies the difficult
issue of stochastic genesis. In addition to the above issue of
super-high energetic genesis of cosmic rays (and associated variable
gamma-ray emissions), the wakefield acceleration mechanism should be a
natural candidate for more compact gamma-ray emission from, e.g., Crab
Nebula \citep{2012ApJ...749...26B,2011Sci...331..739A}, and the
microquasar Cyg.~X-1 \citep{Nowak:2011gs,Ebisuzaki:2014vda}.  With
continued observations of high-energy gamma-ray emission from blazars,
especially those at the highest energies, the underlying emission
processes and acceleration mechanisms may be further revealed.

\section*{Acknowledgements}
NEC and KNA were supported in part by NSF Grants
PHY-1316792, PHY-1620638, and PHY-1915005. TT has been supported by the Norman
Rostoker Fund. We would like to thank fruitful discussions with Aaron
Barth.

\bibliography{/Users/aba/Dropbox/master.bib}

\bibliographystyle{mnras}
\label{lastpage}

\end{document}